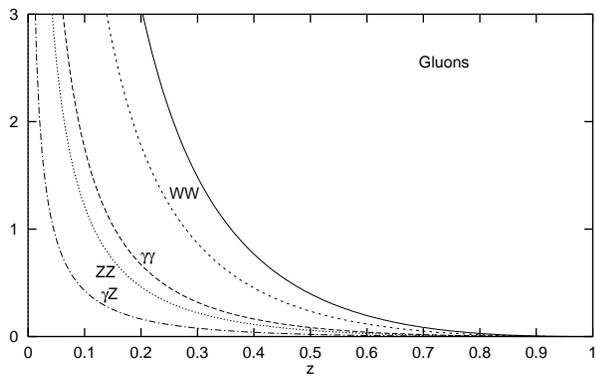

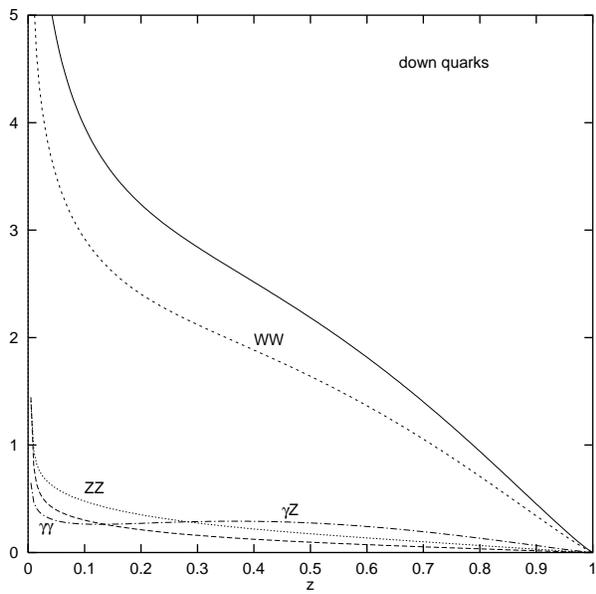

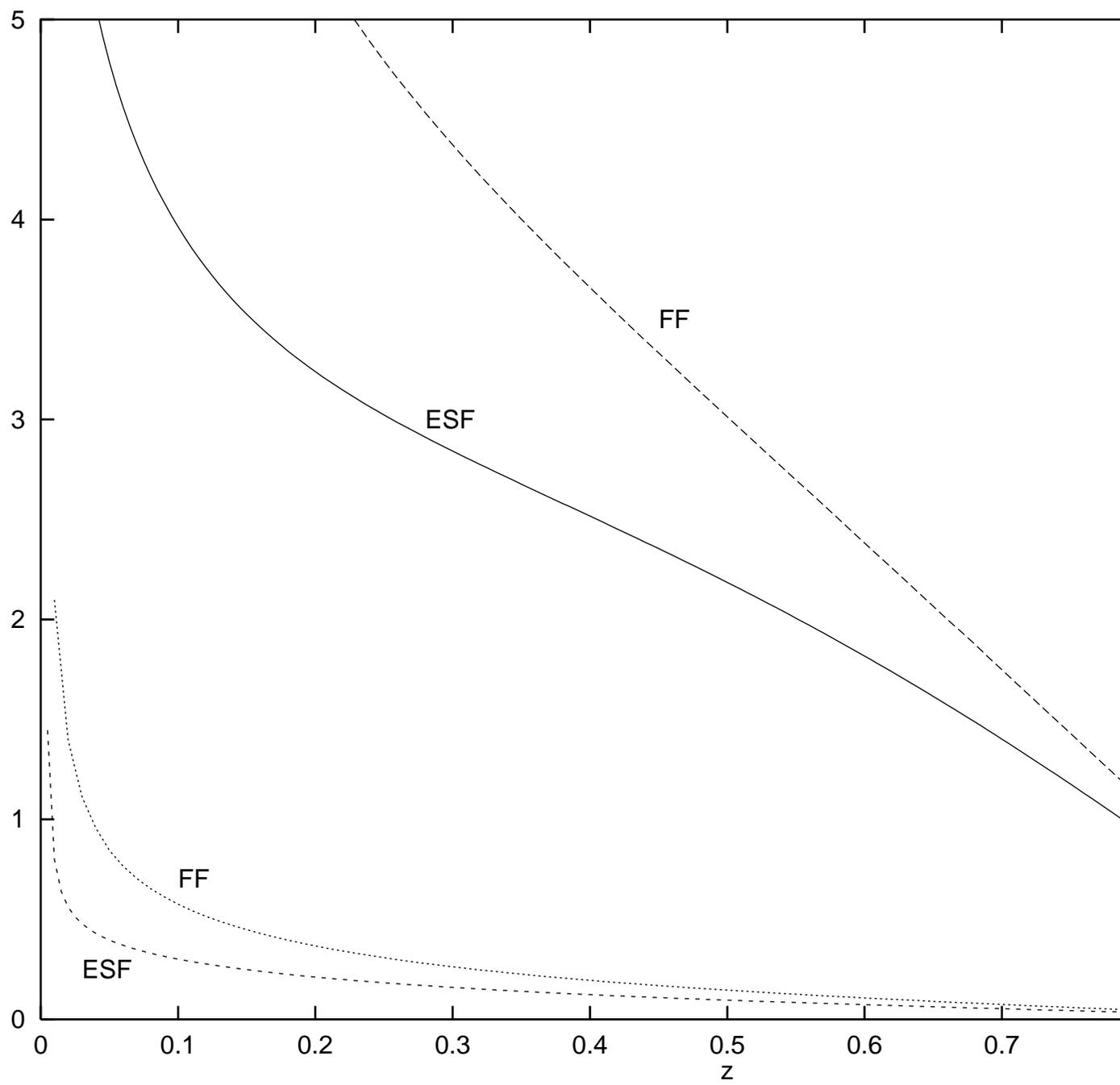

# QCD STRUCTURE OF LEPTONS*

Wojciech SŁOMIŃSKI and Jerzy SZWED†

Physics Department, Brookhaven National Laboratory, Upton, NY 11973, USA
and
Institute of Computer Science, Jagellonian University,
Reymonta 4, 30-059 Kraków, Poland‡

The QCD structure of the electron is defined and calculated. The leading order splitting functions are extracted, showing an important contribution from $\gamma$-$Z$ interference. Leading logarithmic QCD evolution equations are constructed and solved in the asymptotic region where $\log^2$ behaviour of the parton densities is observed. Corrections to the naive evolution procedure are demonstrated. Possible applications with clear manifestation of 'resolved' photon and weak bosons are discussed.

*Work supported by the Polish State Committee for Scientific Research (grant No. 2 P03B 081 09) and the Volkswagen Foundation.
†J. William Fulbright Scholar.
‡Permanent address.

The QCD component of the photon ('resolved' photon) calculated some time ago [1] is now clearly visible in high energy experiments [2]. Theoretical analysis of the quark-gluon component has been recently extended to weak bosons [3] demonstrating new features of the parton densities of 'resolved' W and Z — the calculated densities show strong spin and flavour dependence. In the experiment however one has rarely real weak bosons at one's disposal and in most experiments even the high energy photons are 'nearly on-shell' only. In physical processes, where the structure of electroweak bosons may contribute significantly, it is the lepton, initiating the process which is the source of the intermediate bosons. The standard procedure applied in such cases is to use the equivalent photon approximation [4] (extended also to the case of weak bosons [5]) and, as a next step, to convolute the obtained boson distributions with parton densities inside the bosons. For example, the parton $k$ density inside the electron $F_k^{e^-}$ would read:

$$F_k^{e^-}(z, \hat{Q}^2, P^2) = \sum_B F_B^{e^-}(\hat{Q}^2) \otimes F_k^B(P^2) \tag{1}$$

where $(F \otimes G)(z) \equiv \int dx\, dy\, \delta(z - xy) F(y) G(x)$, $P^2$ is the hard process scale, $\hat{Q}^2$ is the maximum allowed virtuality of the boson (usually taken to be proportional to $P^2$) and $z$ - the momentum fraction of the parton $k$ with respect to the electron (detailed definitions follow).

In such an approximation several questions arise: how far off-shell can the intermediate bosons be, how large are their interference effects, what are the energy scales governing the consecutive steps. It can be easier and more precise to answer the direct question: what is the quark and gluon content of the incoming lepton or, in other words, what is the lepton structure function? In this letter we address the above problem finding several corrections to the standard procedure.

Let us consider inclusive scattering of a virtual gluon off an electron. In the lowest order in the electromagnetic and strong coupling constants ($\alpha$ and $\alpha_s$) the electron couples to $q$-$\bar{q}$ pair as shown in Figure 1. The incoming electron $e$ carries 4-momentum $l$ and the off-shell gluon $G^*$ of 4-momentum $p$ with large $P^2 \equiv -p^2$, serves here as a probe of the electron. In the final state we have a massless quark $q$ and antiquark $\bar{q}$ of 4-momenta $k$ and $k'$ and lepton $\ell'$ (electron or neutrino) of 4-momentum $l'$. The exchanged boson $B = \{\gamma, Z, W\}$ carries 4-momentum $q$ ($Q^2 \equiv -q^2$).

The current matrix element squared for an unpolarized electron reads:

$$\mathcal{J}(e^- \, G^* \to \ell' \, q_\eta \, \bar{q}_{-\eta}) = \mathcal{J}_{\mu\nu}^\eta(l, p) =$$
$$\frac{1}{4\pi} \int d\Gamma_{l'} d\Gamma_k d\Gamma_{k'} (2\pi)^4 \delta_4(k + k' + l' - p - l)$$
$$\times \langle e^- | J_\mu^\dagger(0) | \ell' q_\eta \bar{q}_{-\eta} \rangle \langle \ell' q_\eta \bar{q}_{-\eta} | J_\nu(0) | e^- \rangle, \tag{2}$$

where

$$d\Gamma_k = \frac{d^4 k}{(2\pi)^4} 2\pi \delta(k^2). \tag{3}$$

For massless quarks we can decompose the current in the helicity basis:

$$\mathcal{J}_\sigma^\eta(l, p) = \epsilon_{(\sigma)}^{\mu *}(p) \, \mathcal{J}_{\mu\nu}^\eta(l, p) \, \epsilon_{(\sigma)}^\nu(p), \tag{4}$$

where $\epsilon_{(\sigma)}^\mu(p)$ are polarization vectors of a spin-1 boson with momentum $p^\mu = (p_0, 0, 0, p_z)$:

$$\epsilon_\pm^\mu = \frac{1}{\sqrt{2}}(0, 1, \pm i, 0), \tag{5}$$

$$\epsilon_0^\mu(q) = \sqrt{\frac{1}{|p^2|}}(p_z, 0, 0, p_0). \tag{6}$$

In a frame where $\vec{q}$ is antiparallel to $\vec{p}$ contributions from different helicities of exchanged bosons do not mix and the current reads:

$$\mathcal{J}^\eta_\sigma(p,l) = \frac{\alpha_s \alpha^2}{2\pi} \sum_{A,B,\rho} g^{Aq}_\eta g^{Bq}_\eta \int \frac{dy}{y} P^{e^-}_{A_\rho B_\rho}(y)$$
$$\times \int_{Q^2_{min}}^{Q^2_{max}} \frac{Q^2 dQ^2}{(Q^2 + M^2_A)(Q^2 + M^2_B)} H^\eta_{\rho\sigma}(x, Q^2), \tag{7}$$

where $P^{e^-}_{A_\rho B_\rho}(y)$ describes weak boson emission from the electron and $H^\eta_{\rho\sigma}(x, Q^2)$ — $q$-$\bar{q}$ pair production by virtual gluon and electroweak boson. $g^{Aq}_\eta$ is the boson $A$ to quark $q_\eta$ coupling in the units of proton charge $e$. The sum runs over the electroweak bosons ($A, B = \gamma, W^-, Z$) and their polarizations $\rho = \pm 1, 0$. Note that although the sum is diagonal in the polarization index $\rho$ it is not in the boson type $A, B$. The off-diagonal terms in the sum arisefrom the $\gamma$-$Z$ interference. As demonstrated below their contribution is substantial.

To answer our main problem of 'an electron splitting into a quark' we take the limit $Q^2 \ll P^2$ and keep the leading terms only. Within this approximation the kinematic variables $x, y, z$ read

$$y = \frac{pq}{pl}, \quad z = xy = \frac{-p^2}{2pl}, \tag{8}$$

aquiring the parton model interpretation of the quark momentum fraction ($z$) and of boson momentum fraction ($y$), both with respect to the parent electron. The leading term of the hadronic part does not depend on quark helicity $\eta$:

$$H_{\pm\mp}(x, Q^2) = x^2 \log \frac{P^2}{Q^2}, \tag{9}$$

$$H_{\pm\pm}(x, Q^2) = (1-x)^2 \log \frac{P^2}{Q^2}, \tag{10}$$

with other components finite for $P^2/Q^2 \to 0$.

We also recognize $P^{e^-}_{A_\rho B_\rho}(y)$ as a generalization of the splitting functions of an electron into bosons [3]:

$$P^{e^-}_{A_\rho B_\rho}(y) = \frac{1}{2}(g^{Ae^-}_- g^{Be^-}_- yY_{-\rho} + g^{Ae^-}_+ g^{Be^-}_+ yY_\rho) \tag{11}$$

with

$$Y_+(y) = \frac{1}{y^2}, \tag{12}$$

$$Y_-(y) = \frac{(1-y)^2}{y^2}, \tag{13}$$

$$Y_0(y) = \frac{2(1-y)}{y^2}, \tag{14}$$

where $g^{Ae^-}_\pm$ is the electron to boson $A_\rho$ coupling in the units of proton charge $e$.

From kinematics $y \in [z, 1 - \mathcal{O}(m^2_e/P^2)]$ and the integration limits for $Q^2$ read

$$Q^2_{min} = m^2_e \frac{y^2}{1-y}, \quad Q^2_{max} = P^2 \frac{z + y - zy}{z}, \tag{15}$$

with $m_e$ being the electron mass. Although smaller than the already neglected quark masses, it is the electron mass which must be kept finite in order to regularize colinear divergencies. The upper limit of integration requires particular attention. In general it is a function of $P^2$, however integration up to the maximum kinematically allowed value $Q^2_{\max}$ would violate the condition $Q^2/P^2 \ll 1$. For our approximation to work we integrate over $Q^2$ up to $\hat{Q}^2_{\max} = \epsilon P^2$ where $\epsilon \ll 1$ and generally depends on $y$ and $z$. A similar condition is in fact used in phenomenological applications of the equivalent photon approximation [6]. Integrating Eq.(7) over $Q^2$ within such limits and keeping only leading-logarithmic terms leads to

$$\mathcal{J}^\eta_\sigma(p,l) = \frac{\alpha_s \alpha}{6} \sum_{A,B,\rho} g^{Aq}_\eta g^{Bq}_\eta F^{e^-}_{A_\rho B_\rho}(P^2)$$
$$\otimes [P^\rho_{q_\eta} \delta_{\eta,-\sigma} + P^\rho_{\bar{q}_{-\eta}} \delta_{\eta,\sigma}] \log P^2 \,, \tag{16}$$

where $(F \otimes P)(z) \equiv \int dx\, dy\, \delta(z - xy) F(y) P(x)$.

$P^\rho_{q_\eta}(x)$ and $P^\rho_{\bar{q}_\eta}(x)$ are boson-quark (-antiquark) splitting functions [1,3]

$$P^\pm_{q_\pm}(x) = P^\pm_{\bar{q}_\pm}(x) = 3x^2, \quad P^\mp_{q_\pm}(x) = P^\mp_{\bar{q}_\pm}(x) = 3(1-x)^2 \tag{17}$$

and $F^{e^-}_{A_\rho B_\rho}(y, P^2)$ is the density matrix of polarized bosons inside electron. Its transverse components read

$$F^{e^-}_{\gamma_\pm \gamma_\pm}(y) = \frac{\alpha}{2\pi} \frac{(1-y)^2 + 1}{2y} \log \mu_0 \,, \tag{18a}$$

$$F^{e^-}_{Z_+ Z_+}(y) = \frac{\alpha}{2\pi} \tan^2 \theta_W \frac{\rho_W^2 (1-y)^2 + 1}{2y} \log \mu_Z \,, \tag{18b}$$

$$F^{e^-}_{Z_- Z_-}(y) = \frac{\alpha}{2\pi} \tan^2 \theta_W \frac{\rho_W^2 + (1-y)^2}{2y} \log \mu_Z \,, \tag{18c}$$

$$F^{e^-}_{\gamma_+ Z_+}(y) = \frac{\alpha}{2\pi} \tan \theta_W \frac{\rho_W (1-y)^2 - 1}{2y} \log \mu_Z \,, \tag{18d}$$

$$F^{e^-}_{\gamma_- Z_-}(y) = \frac{\alpha}{2\pi} \tan \theta_W \frac{\rho_W - (1-y)^2}{2y} \log \mu_Z \,, \tag{18e}$$

$$F^{e^-}_{W_+ W_+}(y) = \frac{\alpha}{2\pi} \frac{1}{4 \sin^2 \theta_W} \frac{(1-y)^2}{y} \log \mu_W \,, \tag{18f}$$

$$F^{e^-}_{W_- W_-}(y) = \frac{\alpha}{2\pi} \frac{1}{4 \sin^2 \theta_W} \frac{1}{y} \log \mu_W \,, \tag{18g}$$

where

$$\rho_W = \frac{1}{2 \sin^2 \theta_W} - 1 \,, \tag{19}$$

$\theta_W$ is the Weinberg angle and

$$\log \mu_0 = \log \frac{\epsilon P^2}{m_e^2}, \quad \log \mu_B = \log \frac{\epsilon P^2 + M_B^2}{M_B^2}. \tag{20}$$

All other density matrix elements (containing at least one longitudinal boson) do not develop logarithmic behaviour.

It is natural to introduce at this point the splitting functions of an electron into a quark at the momentum scale $P^2$ as

$$\mathcal{P}_{q_\eta}^{e^-}(P^2) = \sum_{AB} g_\eta^{Aq} g_\eta^{Bq} \sum_\rho F_{A_\rho B_\rho}^{e^-}(P^2) \otimes P_{q_\eta}^\rho \,. \tag{21}$$

The expicit expressions for quarks read

$$\begin{aligned}
\mathcal{P}_{q_+}^{e^-}(z, P^2) =& \frac{3\alpha}{4\pi} \{ e_q^2 \left[ \Phi_+(z) + \Phi_-(z) \right] \log \mu_0 \\
&+ e_q^2 \tan^4 \theta_W \left[ \Phi_+(z) + \rho_W^2 \Phi_-(z) \right] \log \mu_Z \\
&- 2 e_q^2 \tan^2 \theta_W \left[ -\Phi_+(z) + \rho_W \Phi_-(z) \right] \log \mu_Z \} \,,
\end{aligned} \tag{22a}$$

$$\begin{aligned}
\mathcal{P}_{q_-}^{e^-}(z, P^2) =& \frac{3\alpha}{4\pi} \{ e_q^2 \left[ \Phi_+(z) + \Phi_-(z) \right] \log \mu_0 \\
&+ z_q^2 \tan^4 \theta_W \left[ \Phi_-(z) + \rho_W^2 \Phi_+(z) \right] \log \mu_Z \\
&+ 2 e_q z_q \tan^2 \theta_W \left[ -\Phi_-(z) + \rho_W \Phi_+(z) \right] \log \mu_Z \\
&+ (1 + \rho_W)^2 \Phi_+(z) \delta_{qd} \log \mu_W \} \,,
\end{aligned} \tag{22b}$$

where

$$\Phi_+(z) = \frac{1-z}{3z} (2 + 11z + 2z^2) + 2(1+z) \log z \,, \tag{23}$$

$$\Phi_-(z) = \frac{2(1-z)^3}{3z} \,, \tag{24}$$

and

$$z_q = \frac{T_3^q}{\sin^2 \theta_W} - e_q \,, \tag{25}$$

with $e_q$ and $T_3^q$ being the quark charge and 3-rd weak isospin component, respectively. The splitting functions for antiquark of opposite helicity can be obtained from Eq.(22) by interchanging $\Phi_+$ with $\Phi_-$.

The splitting functions introduced above show two new features. The first one, already mentioned before, is the contribution from the interference of electroweak bosons ($\gamma$ and $Z$ only). The second is their $P^2$ dependence, which arises from the upper integration limit $\hat{Q}_{\max}^2$.

The above splitting functions, when cast into the evolution equations, produce nontrivial effects in the $P^2$-dependence. We consider the evolution equations in first order in electroweak couplings and leading logarithmic in QCD. Introducing $t = \log(P^2/\Lambda_{\text{QCD}}^2)$ one can write the standard evolution equations for the density of polarized QCD partons $i_\eta$ (quarks, antiquarks and gluons):

$$\begin{aligned}
\frac{dF_{i_\eta}^{e^-}(t)}{dt} =& \frac{\alpha}{2\pi} \sum_{B,\rho} P_{i_\eta}^{B_\rho} \otimes F_{B_\rho}^{e^-}(t) \\
&+ \frac{\alpha_s(t)}{2\pi} \sum_{k,\rho} P_{i_\eta}^{k_\rho} \otimes F_{k_\rho}^{e^-}(t) \,,
\end{aligned} \tag{26}$$

where $P_{i_\eta}^{k_\rho}$ are the standard QCD parton splitting functions [8]. The first sum runs over electroweak bosons while the second one — over QCD partons. We immediately recognize Eq.(21) to be the generalization of the first term in the above equation. Remembering that there is no direct coupling of the electroweak sector to gluons we arrive at the master equations for the parton densities inside the electron:

$$\frac{dF_{q_\eta}^{e^-}(t)}{dt} = \frac{\alpha}{2\pi}\mathcal{P}_{q_\eta}^{e^-}(t) + \frac{\alpha_s(t)}{2\pi}\sum_{k,\rho} P_{q_\eta}^{k_\rho} \otimes F_{k_\rho}^{e^-}(t) \tag{27a}$$

$$\frac{dF_{G_\lambda}^{e^-}(t)}{dt} = \frac{\alpha_s(t)}{2\pi}\sum_{k,\rho} P_{G_\lambda}^{k_\rho} \otimes F_{k_\rho}^{e^-}(t). \tag{27b}$$

We stress that the convolution of the equivalent boson distributions and boson-quark splitting functions, Eq.(21), occurs at the level of splitting functions. It is not equivalent to the usually performed convolution of the distribution functions Eq.(1) because of the $P^2$ dependence of the boson distribution functions Eq.(18). Only in the case when the upper limit of integration $\hat{Q}_{\max}^2$ is kept fixed ($P^2$-independent), e.g. by special experimental cuts, are the convolutions equivalent at both levels.

The equations Eq.(27) can be solved in the asymptotic $t$ region where we approximate the strong coupling constant

$$\alpha_s(t) \simeq \frac{2\pi}{bt}, \tag{28}$$

with $b = 11/2 - n_f/3$ for $n_f$ flavours. The asymptotic (large $t$) solution to Eqs.(27) for the parton $k$ of polarization $\rho$ can be now parametrized as

$$F_{k_\rho}^{e^-}(z,t) \simeq \frac{1}{2}\left(\frac{\alpha}{2\pi}\right)^2 f_{k_\rho}^{\mathrm{as}}(z)\, t^2 \tag{29}$$

resulting in purely integral equations

$$f_{i_\rho}^{\mathrm{as}} = \hat{\mathcal{P}}_{i_\rho}^{e^-} + \frac{1}{2b}\sum_{k,\lambda} P_{i_\rho}^{k_\lambda} \otimes f_{k_\lambda}^{\mathrm{as}}, \tag{30}$$

where $\hat{\mathcal{P}}_{i_\rho}^{e^-}(z)$ are given by Eqs.(22) with all $\log \mu_A \equiv 1$.

Numerical solutions to the above equations, with the use of method described in Ref. [3], are presented in Figure 2 for the unpolarized quark and gluon distributions. One notices significant contribution from the $W$ intermediate state in the d-type quark density (this would be even more pronounced if we looked at the left-handed d quarks). The most surprising however is the $\gamma$-$Z$ interference contribution which cannot be neglected, as it is comparable to the $Z$ term. It violates the standard probabilistic approach where only diagonal terms are taken into account. This also stresses the necessity of introducing the concept of electron structure function in which all contributions from intermediate bosons are properly summed up. The asymptotic solutions to the polarized parton densities $\Delta q = q_+ - q_-$ and $\Delta G = G_+ - G_-$ are given in Ref. [7]. Due to the nature of weak couplings they turn out to be nonzero, even in the case of gluon distributions. Again the $\gamma$-$Z$ interference term is important and the $W$ contribution dominates in the asymptotic region.

One should keep in mind that at finite $t$ the logarithms multiplying the photon contribution differ from the remaining ones (Eq.(20)). Being scaled by $m_e$, they lead to the photon domination at presently available $P^2$. The importance of the interference term remains constant relative to the $Z$ contribution, as they are both governed by the same logarithm. But even at presently available momenta, where the 'resolved' photon dominates, one can see how the correct treatment of the scales changes the evolution. In Fig. 3 we present the asymptotic solutions of the evolution equations following from our procedure (ESF) compared to those following from naive application of the convolution Eq.(1) (FF). It is possible that the difference can be traced in the analysis of presently available data. This question is currently under study.

To summarize we have presented a construction of the electron structure functions within leading logarihtmic approximation to QCD and leading order in electroweak interactions. The $\gamma$-$Z$ interference, contrary to naive expectations, turns out to be important. Direct calculation of the splitting functions of an electron into quarks allows for precise control of the momentum scales entering the evolution. It also shows that the convolution of leptons, electroweak bosons and quarks should be made at the level of splitting functions rather than distribution functions. Unless forced otherwise by the experimental cuts, the electron splitting functions depend on the external scale $P^2$ and influence significantly the parton evolution. Phenomenological applications of the above analysis require very high momentum scales in order to see the weak boson and interference contributions. Possible processes where these effects could show up include heavy flavour, large $p_\perp$ jet and Higgs boson production in lepton induced processes. At lower momenta, where the photons dominate, the use of the electron structure function allows to treat correctly the parton evolution.

**Acknowledgements.**

The authors would like to thank the Theory Groups of Brookhaven National Laboratory and DESY for their hospitality.

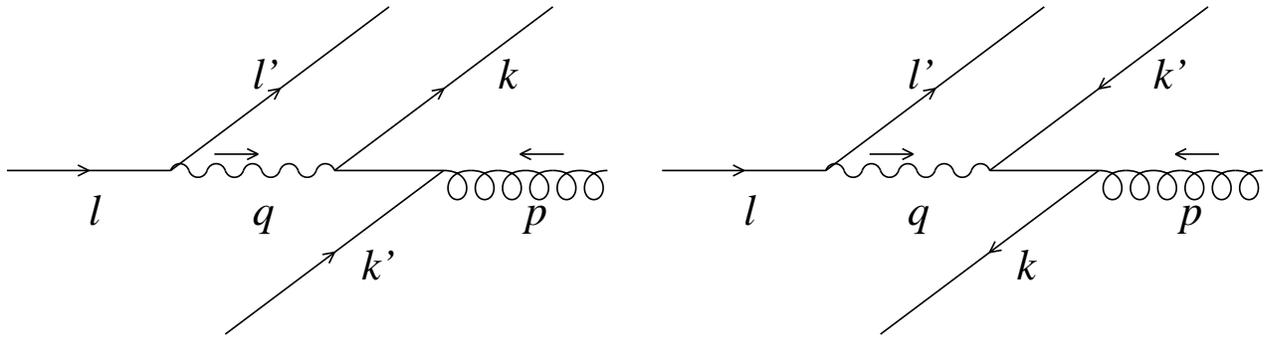

FIG. 1. Lowest order graphs contributing to the process: $e + G^* \to \ell' + q + \bar{q}$.

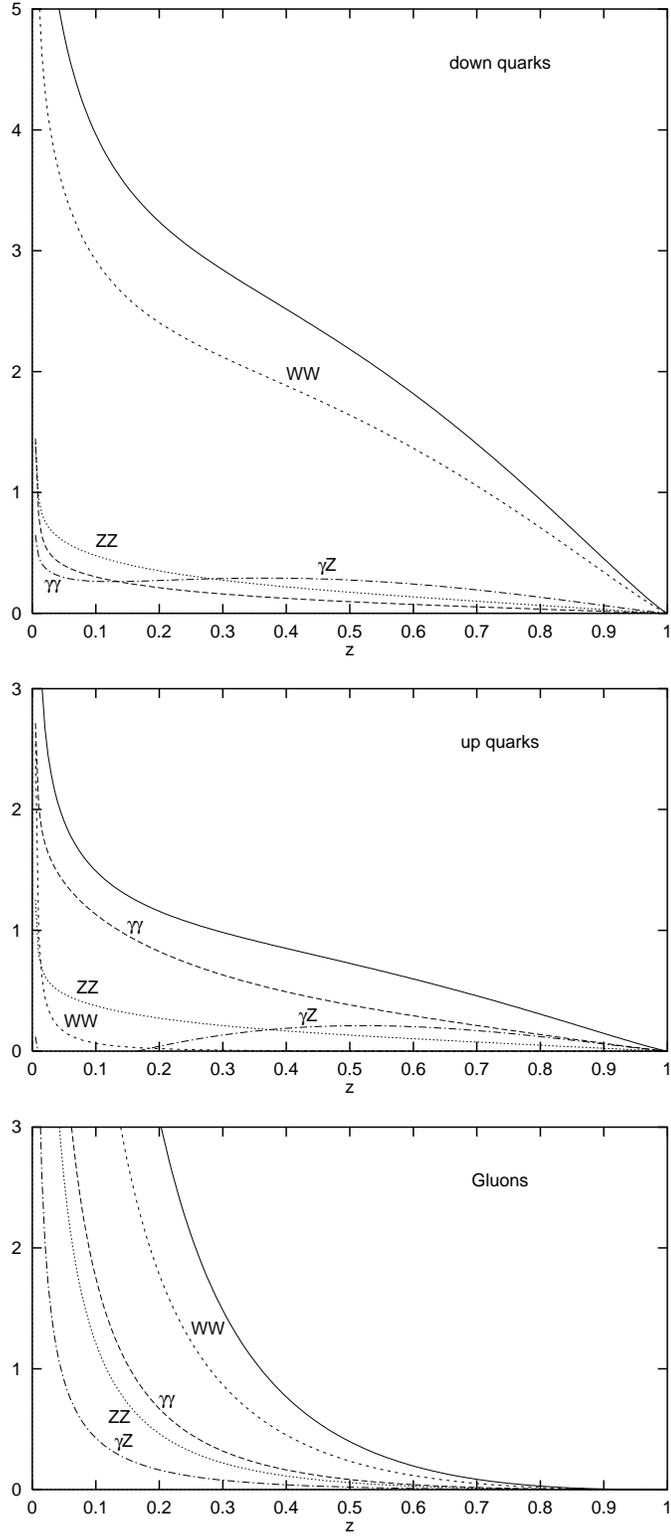

FIG. 2. Unpolarized quark and gluon distributions $zf^{as}(z)$ — solid line. The other lines show contributions from different electroweak bosons.

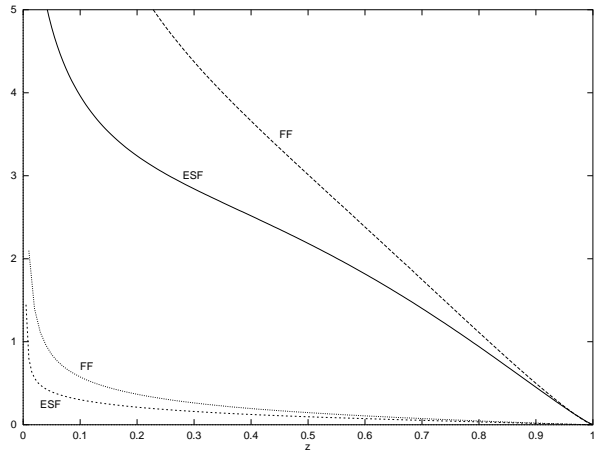

FIG. 3. Comparison of unpolarized d-quark distributions $zf^{as}(z)$ calculated by ESF and FF methods. The upper two lines result from contributions from all electroweak bosons while the lower two from $\gamma\gamma$ only.

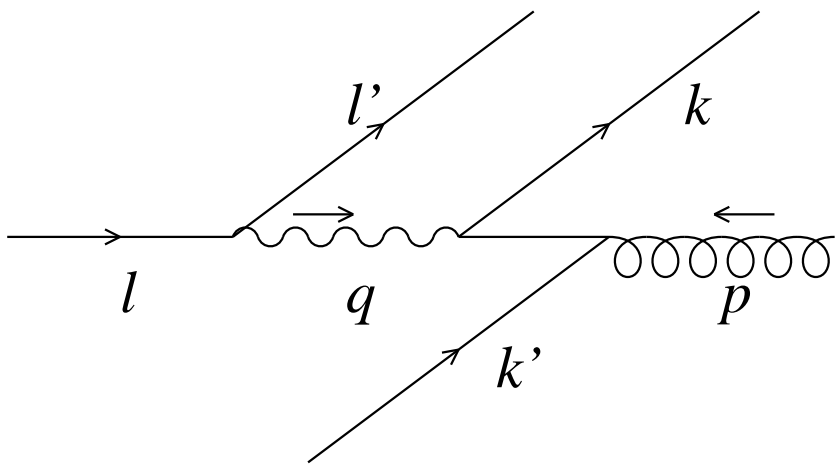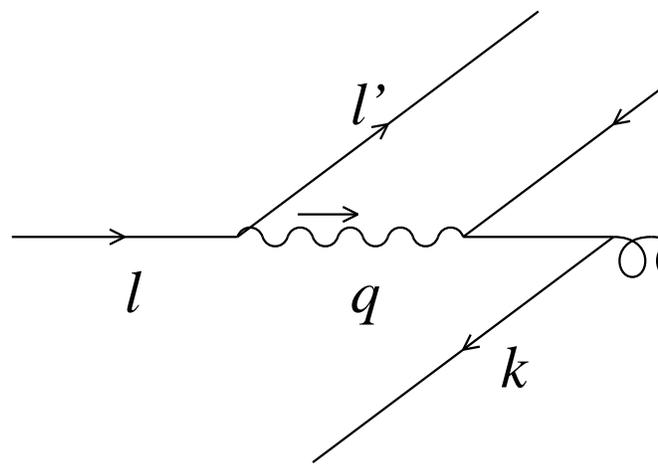

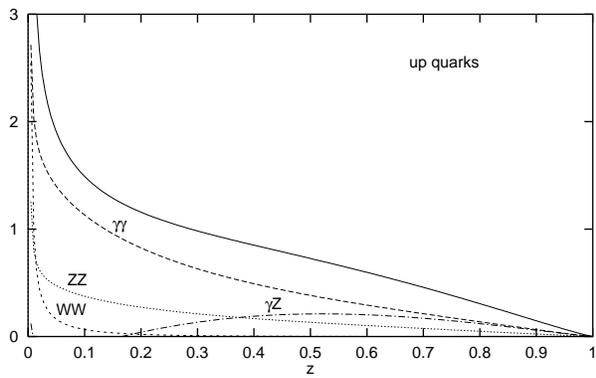